\documentstyle[12pt,epsf]{article}

\newfont{\lfont}{line10}

\newcommand{\BE}{\begin{equation}}
\newcommand{\EE}{\end{equation}}
\newcommand{\BEA}{\begin{eqnarray}}
\newcommand{\EEA}{\end{eqnarray}}

\def\12{\frac{1}{2}}
\def\bea{\begin{eqnarray}}
\def\eea{\end{eqnarray}}
\def\ba{\begin{array}}
\def\ea{\end{array}}

\def\one-loop{\mbox{\scriptsize one-loop}}

\def\G{\Gamma}

\catcode`\@=11

\@addtoreset{equation}{section}
\def\theequation{\arabic{section}.\arabic{equation}}

\def\@normalsize{\@setsize\normalsize{15pt}\xiipt\@xiipt
\abovedisplayskip 14pt plus3pt minus3pt%
\belowdisplayskip \abovedisplayskip
\abovedisplayshortskip  \z@ plus3pt%
\belowdisplayshortskip  7pt plus3.5pt minus0pt}

\def\small{\@setsize\small{13.6pt}\xipt\@xipt
\abovedisplayskip 13pt plus3pt minus3pt%
\belowdisplayskip \abovedisplayskip
\abovedisplayshortskip  \z@ plus3pt%
\belowdisplayshortskip  7pt plus3.5pt minus0pt
\def\@listi{\parsep 4.5pt plus 2pt minus 1pt
              \itemsep \parsep
              \topsep 9pt plus 3pt minus 3pt}}

\def\underline#1{\relax\ifmmode\@@underline#1\else
          $\@@underline{\hbox{#1}}$\relax\fi}
\@twosidetrue

\relax

\catcode`@=12

\catcode`\@=11

\def\section{\@startsection{section}{1}{\z@}{3.5ex plus 1ex minus
     .2ex}{2.3ex plus .2ex}{\large\bf}}

\def\thesection{\Roman{section}.}

\def\appendix{\setcounter{section}{0}
          \def\thesection{Appendix }
         \def\theequation{\Alph{section}.\arabic{equation}}}

\def\ps@headings{\def\@oddfoot{}\def\@evenfoot{}
\def\@oddhead{\hbox{}\hfill
          \makebox[.5\textwidth]{\raggedright\ignorespaces --\thepage{}--
          \hfill {}}}
\def\@oddhead{\hbox{}\hfill --\thepage{}-- \hfill
          {}}
\def\@evenhead{\@oddhead}
\def\subsectionmark##1{\markboth{##1}{}}
}

\ps@headings

\catcode`\@=12

\relax

\def\figcap{\section*{Figure Captions\markboth
          {FIGURECAPTIONS}{FIGURECAPTIONS}}\list
          {Fig. \arabic{enumi}:\hfill}{\settowidth\labelwidth{Fig. 999:}
          \leftmargin\labelwidth
          \advance\leftmargin\labelsep\usecounter{enumi}}}
 \relax
\def\tablecap{\section*{Table Captions\markboth
          {TABLECAPTIONS}{TABLECAPTIONS}}\list
          {Table \arabic{enumi}:\hfill}{\settowidth\labelwidth{Table 999:}
          \leftmargin\labelwidth
          \advance\leftmargin\labelsep\usecounter{enumi}}}
 \relax
\def\reflist{\section*{References\markboth
          {REFLIST}{REFLIST}}\list
          {[\arabic{enumi}]\hfill}{\settowidth\labelwidth{[999]}
          \leftmargin\labelwidth
          \advance\leftmargin\labelsep\usecounter{enumi}}}
 \relax

\catcode`\@=11

\def\ps@headings{\def\@oddfoot{}\def\@evenfoot{}
\def\@oddhead{\hbox{}\hfill
          \makebox[.5\textwidth]{\raggedright\ignorespaces --\thepage{}--
          \hfill {}}}
\def\@evenhead{\@oddhead}
\def\subsectionmark##1{\markboth{##1}{}}
}

\ps@headings

\relax

\newskip\humongous \humongous=0pt plus 1000pt minus 1000pt

\newif\ifdtup

\def\beq{\begin{equation}}
\def\eeq{\end{equation}}

\def\beqn{\begin{eqnarray}}
\def\eeqn{\end{eqnarray}}
\relax

\def\G2{{\; \rm GeV/}c2}
\def\G{\; \rm GeV}

\def\dotx{\dotx{\dot\overline{x}}}

\relax

\begin{document}
%
%
\begin{titlepage}

\renewcommand{\thefootnote}{\fnsymbol{footnote}}

\begin{flushright}
        \normalsize
      December, 2003 \\
    OCU-PHYS 206 \\
    hep-th/0312306  \\

\end{flushright}

%
\begin{center}
    {\large\bf Whitham Prepotential and Superpotential }
\end{center}

\vfill

\begin{center}
      {%
H. Itoyama\footnote{e-mail: itoyama@sci.osaka-cu.ac.jp}
\quad and \quad
H. Kanno\footnote{e-mail: kanno@math.nagoya-u.ac.jp}
}\\
\end{center}

\vfill

\begin{center}
        ${}^{\dag}$\it  Department of Mathematics and Physics,
          Graduate School of Science\\
          Osaka City University\\
          3-3-138, Sugimoto, Sumiyoshi-ku, Osaka, 558-8585, Japan  \\
\vskip 1.0em
        ${}^{\ddag}$\it Graduate School of Mathematics, Nagoya University \\
         Chikusa-ku, Nagoya, 464-8602, Japan  \\
\end{center}

\vfill

\begin{abstract}
  ${\cal N}=2$ supersymmetric
  $U(N)$ Yang-Mills theory softly broken to ${\cal N}=1$ by 
 the superpotential of  the adjoint scalar fields  
 is discussed from the viewpoint of the Whitham deformation
  theory for prepotential.  
 With proper identification of the superpotential
 we derive the matrix model curve from
 the condition that the mixed second derivatives of the Whitham
 prepotential have a nontrivial kernel. 

\end{abstract}

\vfill

\setcounter{footnote}{0}
\renewcommand{\thefootnote}{\arabic{footnote}}

\end{titlepage}
\section{Introduction}

Gauge theory with ${\cal N}=2$ supersymmetry
  has a successful exact description of low energy
  effective action in terms of
  curve, the meromorphic differential,
 the period integrals and above all the
 prepotential \cite{SW,SWC}.  This structure is shared by
  the corresponding classical integrable system of particles,
  which may be regarded as representing universality class of
 the field theories of this type \cite{SWINT}. 
In particular, the prepotential,
  which extracts all available information on low energy phenomena
  from the curve and the periods,  can be extended to include
 infinite number of couplings as time variables
\cite{IM2, GMMM, Edel}  and becomes  a generating functional
 (or $\tau$ function) of some kind
  from the point of view of the integrable hierarchy. 
 It is therefore natural to investigate what role is played by
  the prepotential when the original theory is finitely
  as opposed to infinitesimally driven
 by  a specific set of operators.
The framework which fits this line of thoughts is
  deformation theory of prepotential
 called Whitham deformation. See \cite{whithamrev} for reviews.
 
In more physical terms, this amounts to asking whether
   the notion of the prepotential be effective
   when the ${\cal N}=2$ theory becomes softly broken to
  ${\cal N}=1$  \cite{DS} or ${\cal N}=0$.
 In fact there have been several such developments till now,
 a few of which are listed below.
 In \cite{Edel}, the Whitham deformation based on the
 formalism of \cite{IM2} and the computation of \cite{GMMM}
 has been applied to ${\cal N}=0$ through  the spurion
 formalism of \cite{LAG}.
  More recently, the gluino condensate prepotential has 
 appeared through the determination of the effective superpotential
for ${\cal N}=1$ super Yang-Mills with the adjoint chiral superfield
\cite{CIVCVTV}.  This prepotential has been identified as the
 matrix model free energy \cite{DV}.  
In the subsequent development of
 \cite{CDSW}, the prepotential appeared via 
a fermionic shift symmetry \footnote{The developments concerning
  with this
 particular point include the work of \cite{ACDGN,IK1}.  }
 associated with  the system of three anomalous chiral
 Ward-Takahashi identities \cite{Konishi}. 
 See a recent review \cite{Argreview} for more on this last
 development,
 and \cite{DebHollo} for  recent discussions on the
  relationship of the superpotential with the Lax matrix.

In this paper, we will add one more step to this list of developments
 in prepotential theory:
  we will find a more direct route and rationale leading to
 the appearance of  the prepotential for ${\cal N}=1$ theories.
 We begin with a few preliminaries.  In the next section,
 we first give the basic ingredients of the Seiberg-Witten curve
 for the  ${\cal N}=2$ $SU(N)$ pure-super Yang-Mills theory
(or the spectral curve of the $N$ site periodic Toda chain.)
 Next, we briefly review Whitham deformation theory for the
 prepotential. 
It is a generic framework and  we need  to find a proper condition
 that deforms the theory and that breaks ${\cal N}=2$
 supersymmetry to ${\cal N}=1$. At the same time, 
 the superpotential  responsible for this deformation
 must be identified.
In section three, we carry out this task.
We examine the mixed second derivatives.
  A main thrust of this paper is to observe that, in  prepotential
  theory, the condition for a curve  to degenerate or factorize is
given by that the kernel of 
 the matrix  made of the mixed second derivatives of the
 Whitham  prepotential  be nontrivial:
\beqn
 {\rm ker}  \frac{ \partial^{2} \cal{F}}{\partial a^{i} T_{\ell}}
  \neq {\bf 0} \;\;.
\eeqn
We derive the matrix model curve,
 identifying the tree-level superpotential with 
the vector belonging to this nontrivial kernel of
 the matrix above.
  In the classical limit where the scale parameter vanishes, 
  expressions get simplified largely. We discuss this case
 separately in section four and show that the identification
  made in section three is indeed correct.
  We make several comments on the gluino condensate prepotential
  in the final section.

\section{Some Preliminaries}
\subsection{ The Seiberg-Witten curve}
Let us first collect some basic ingredients of
 the Seiberg-Witten curve  for the
  ${\cal N}=2$ $SU(N)$ pure-super Yang-Mills theory (or the spectral curve of the $N$ site periodic Toda chain)
 and the meromorphic differential $dS_{SW}$ defined on this Riemann
 surface of genus $g= N-1$.
The curve can be represented by a hyperelliptic form:
\beqn
\label{hyel}
  Y^{2} &=& P_{N}(x)^{2} - 4 \Lambda^{2N} \;\;,   \\
  P_{N}(x) &\equiv& \langle \det \left(x {\bf 1} - \Phi \right)
   \rangle = \prod_{i=1}^{N} \left( x- p_{i} \right)
    = x^{N} - \sum_{k=2}^{N} u_{k} x^{N-k} \;\;  \nonumber \\
  &=&  \sum_{k=0}^{N} {\it s}_{k}(h_{\ell})  x^{N-k} \;\;,
\eeqn
 where ${\it s}_{k}(h_{\ell})$  are
 the fundamental Schur polynomials of
\beqn
 h_{\ell} = \frac{1}{\ell} \langle tr_{N} \Phi^{\ell} \rangle
  = \frac{1}{\ell} \sum_{i=1}^{N} p_{i}^{\ell} \;,  \;\; \ell = 2,
 \cdots N \;\;.
\eeqn
 The overall $U(1)$  can be decoupled from our consideration
  at this moment.
In the Toda chain representation, this curve is
 parameterized by the spectral parameter $z$ by
\beqn
\label{zrep}
  P_{N}(x) &=& z +  \frac{\Lambda^{2N}}{z} \;\;,  \nonumber \\
 Y &=& z -  \frac{\Lambda^{2N}}{z} \;\;.
\eeqn
 The general variations of eq. (\ref{zrep}) give us the following
 equations:
\beqn
\label{variation}
   P_{N}^{\prime}  \delta x +   \delta_{u} P_{N} =
       Y \delta \log z  - N 
\left( Y- P_{N} \right)  \delta \log \Lambda   \;\;,  \nonumber \\
 \delta Y  =   P_{N} \delta \log z  + N 
 \left( Y- P_{N} \right)  \delta \log \Lambda\  \;\;. 
\eeqn
Here $\delta_{u}$ denotes a generic variation with respect to $u_{k}$.
 The distinguished meromorphic differential for the prepotential
 theory of the  ${\cal N}=2$ $SU(N)$ pure-super Yang-Mills is
\beqn
dS_{SW}  &=& x d \log z
= x t(x) dx \;\;. \\
 t(x) &=& \frac{P^{\prime}_{N}}{ \sqrt{P_{N}^{2} - 4 \Lambda^{2N}} }
  \;\;.
\eeqn 
 This differential possesses a double pole at $x= \infty_{\pm}$.
 Moduli derivatives with either $z$ or $x$ and $\Lambda$ fixed
 generate  the bases of the holomorphic differentials: 
\beq
\label{dvk}
 \frac{\partial}{\partial u_{k}}  dS_{SW} \mid_{z, \Lambda}
= \frac{x^{N-k}}{Y} dx  \equiv  dv_{k} \;\;, 
 \;\;     \frac{\partial}{\partial u_{k}}  dS_{SW} \mid_{x, \Lambda}
=   dv_{k} - d \left( \frac{x^{N-k+1}}{Y} \right) \;\;
,  \;\; k= 2, \cdots N \;\;.
\eeq
The differentials $dv_{k}$ are related  by $u_{k}$ dependent
 numerical factors to the canonical holomorphic
 differentials $d \omega_{i}$
which are  normalized  as $\oint_{A_{i}}d \omega_{j}= \delta^{i}_{j}$.
 Moduli derivatives are in general coordinate dependent and
 in this paper, we take them, keeping $z$ fixed.

\subsection{Whitham Prepotential} 

It is possible to deform consistently both moduli of the Riemann
 surface  and the meromorphic differential discussed in the last subsection without losing their defining properties.
Namely,
\beqn
 dS_{SW} \longrightarrow dS  \;\;, \;\;
\frac{\partial}{\partial u_{k}} dS  \mid_{z, \Lambda}
  =  {\rm holomorphic} \;\;.
\eeqn
 This is called Whitham deformation (of Toda integrable hierarchy).
 It has an effect of adding  higher order poles
  to the original double poles of the meromorphic differential
  $dS_{SW}$.
 The deformation is  characterized by a set of the punctures and
  local coordinates in their neighborhood, which we denote
 generically by $\xi$.
In order to accomplish this deformation, we introduce a set of
 meromorphic differentials  $d \Omega_{\ell}$ which satisfy
\beqn
 d \Omega_{\ell} =  \xi^{-\ell -1} d\xi + {\rm nonsingular \; part} 
  \;\;, \;\;  \ell = 1, 2 \cdots \;\;.
\eeqn
This still leaves us with the possibility to add any linear combination
 of the holomorphic differentials $d \omega_{i}$
 to the right hand side.
To remove this ambiguity,  we require the condition
\beq
\oint_{A_{i}} d \Omega_{\ell} =0 \;\;.
\eeq

  Construction of the Whitham prepotential  begins with introducing
   time variables  via
\beqn
 \frac{ \partial dS}{ \partial T_{\ell}}
  =  d \Omega_{\ell} \;\;.
\eeqn
Let
\beq
  a^{i} \equiv \oint_{A_{i}} dS  \;\;, \;\;
\eeq
be  the local coordinates of the moduli space.
  After some reasonings \cite{IM2}, one concludes
\beqn
\label{dS}
 dS = \sum_{i=1}^{g} a^{i} d\omega_{i}
+ \sum_{\ell} T_{\ell} d\Omega_{\ell} \;\;,
\eeqn
 and
\beq
 T_{\ell} = res _{\xi =0} \xi^{\ell} dS  \;.
\eeq
 Both $a^{i}$ and  $T_{\ell}$ are regarded as independent variables.
 In $dS_{SW}$, the first time alone is turned on. 
Invariant moduli $h_{k}$ are expressible as
\beq
 h_{k} =  h_{k}( a^{i}, T_{\ell}) \;\;.
\eeq
The prepotential 
${\cal F}\left( a^{i}, T_{\ell}\right)$
  is introduced via 
\beqn
\frac{\partial {\cal F}}{\partial a^{i}} &=& \oint_{B^{i}}
 d S \;, \;\; \;
  1 \leq i \leq N-1   \label{a1srpre}   \\ 
\frac{\partial {\cal F}}{\partial T_{\ell}} &=& \frac{1}{2\pi i \ell}
  res \xi^{-\ell} dS  \equiv {\cal H}_{\ell}
 \left( h_{k} \right) \;.  \;\;\;  \ell = 1, 2, \cdots
 \label{T1srpre}   
\eeqn
The consistency of eq. (\ref{a1srpre}) and that of eq. (\ref{T1srpre})
 are ensured
  respectively by the property
 that the period matrix is symmetric  as well as by the
  Riemann identity. The right hand side of eq. 
(\ref{T1srpre}),
 which we have denoted by ${\cal H}_{\ell} \left( h_{k} \right)$,
 is some polynomial \cite{GMMM} of the invariant moduli.
  Below, the local coordinates  in the neighborhood of
 the punctures at $x= \infty_{\pm}$ are taken as
\beq
 \xi = z^{\mp \frac{1}{N}} \;\;.
\eeq

\section{Mixed Second Prepotential Derivatives: Derivation
 of the Matrix Model Curve}

 Now we come to the point of our paper.
 Differentiating eqs. (\ref{a1srpre}) and (\ref{T1srpre})
once again, we obtain two distinct formulas
for the mixed second derivatives:
\beqn
\label{mixed}
\frac{\partial^{2} {\cal F}}{\partial a^{j} \partial T_{\ell}}
 =  \oint_{B^{j}} d \Omega_{\ell} 
 = \frac{1}{2\pi i \ell}
  res_{\xi =0} \xi^{-\ell} d \omega_{j} \;\;,
\eeqn
  where $j= 1 \sim N-1$. As for $\ell$,  it is any positive integer.
 Note that eq. (\ref{mixed}) has been given with no reference to $dS$.
The discussion in what follows holds, therefore, for 
the  Whitham prepotential deformed by the arbitrary
  values of the time variables.
 As is stated in the introduction, we impose the condition that
 the kernel of this rectangular matrix be nontrivial:
\beqn
\label{kernel}
 {\rm ker}  \frac{ \partial^{2} \cal{F}}{\partial a^{i} T_{\ell}}
  \neq {\bf 0} \;\;.
\eeqn
Equivalently
\beq
 {\rm rank} \; \frac{ \partial^{2} \cal{F}}{\partial a^{i} T_{\ell}}
  \leq N-2 \;\;.
\eeq
In the case in which the time variables are truncated to
$T_{1}, T_{2}  \cdots T_{N-1}$,  the condition is of course  stated as
that of the vanishing determinant:
\beqn
\label{vandet}
 \det  \frac{ \partial^{2} \cal{F}}{\partial a^{i} T_{\ell}}
  = 0\;\;.
\eeqn

   Eq. (\ref{kernel}) tells us the existence of at least one
 nonvanishing  column vector
\beqn
 \left(
\begin{array}{c}
   c^{1}   \\
   \vdots  \\
   c^{N-1} \\
   \vdots
\end{array}
 \right)
\eeqn   
 belonging to the kernel; 
\beqn
 0=  \sum_{\ell}
 \frac{ \partial^{2} \cal{F}}{\partial a^{i} T_{\ell}} c^{\ell}
  &=& \sum_{\ell}  \oint_{B^{i}} c^{\ell} d \Omega_{\ell} 
   \;\; \label{st1}   \\  
  &=&   \frac{1}{2\pi i} res_{\xi = 0} \left( \sum_{\ell}
  \frac{c_{\ell}}{\ell} \xi^{-\ell}  d \omega_{i} \right)  \;\;.
  \label{st11}
\eeqn

  Eq. (\ref{st1}) implies  the existence 
 of the meromorphic one-form
\beq
  d \tilde{\Omega} \equiv   \sum_{\ell} 
   c^{\ell} d \Omega_{\ell}  \;\;. 
\eeq
 whose period integral over any of the $A_{i}$ and $B^{i}$
 cycles vanishes.
 Let us argue implications of this last statement.
 Obviously, once this property holds, we can integrate 
$d  \tilde{\Omega}$ along any path ending with a point $z$ to define
  a function holomorphic except at punctures: 
$f(z) = \int^{z} d  \tilde{\Omega}$. As for the order of the poles at
 the punctures, it is generically arbitrary according to the
 construction.  This is, however, contradictory to the Weierstrass gap
 theorem  \cite{RSFKRA}, 
which says that,  for $g= N-1$ integers satisfying 
$1 = n_{1} < n_{2} < \cdots < n_{g} < 2g$,
such function  with a pole of order $n_{j}$ does not exist.
This theorem is derived from the Riemann-Roch theorem.
 In order to avoid the contradiction, the degeneration of the
 surface must take place.

 Eq. (\ref{kernel}) also implies  the existence of at least one
 non-vanishing co-vector (row vector)
\beq
 \left( \tilde{c}_{1}, \tilde{c}_{2}, \cdots, \tilde{c}_{N-1} \right)
\eeq
such that
\beq
 0=
\sum_{i =1}^{N-1} \tilde{c}_{i}
 \frac{ \partial^{2} \cal{F}}{\partial a^{i} T_{\ell}} 
  = \sum_{i =1}^{N-1} \tilde{c}_{i}
 \frac{ \partial  {\cal H}_{\ell +1} }{\partial a^{i}} 
  \;\;.  
\eeq
  We see from eq. (\ref{T1srpre}) that this is satisfied provided
\beq
\label{st2}
\sum_{i =1}^{N-1} \tilde{c}_{i}
 \frac{ \partial  {h}_{k} }{\partial a^{i}} = 0 \;\;,
 \;\;  k = 2 \sim N.  
\eeq
 Eq. (\ref{st2}) tells us the degeneration of the curve from the
 point of view of the moduli space: the invariant moduli
 $h_{k}\;, \; k= 2\sim N$   become functions which
 actually depend on less than $N-1$ arguments. 
  This equation is regarded as the counterpart of the statement of
 the vanishing discriminant \cite{discriminants}.

We conclude that the original curve must get degenerated under
  the condition eq. (\ref{kernel}) and
 the factorization of eq. (\ref{hyel}) by the following kind \cite{CIVCVTV} takes place:
\beqn
 Y^{2} (x) &=&  H_{N-n}^{2}(x) F_{2n}(x) \;\;.    \\
  P_{N}^{\prime}(x) &=& H_{N-n}(x) R_{n-1}(x)  \\
  t(x) &=&  \frac{  R_{n-1}(x)}{ \sqrt{F_{2n}(x)}} \;\;.
\eeqn
 The polynomial $R_{n-1}$ does not depend on $N$ \cite{Gopa}.
As is well-known, $N-n$ non-intersecting cycles vanish once this
condition is satisfied and  the moduli space of the resulting
 vacua is  codimension $N-n$ subspace of the original
${\cal N}=2$ Coulomb branch.
  To smoothen out the singularities developed, let us imagine
  blowing up the surface and multiplying  $dv_{k}$ in eq. (\ref{dvk})
   by a factor which cancels the poles developed and which  behaves
  as $x$ at the infinities.  Upon partial fractions, $n$ independent
 differentials emerge from $dv_{k}, k= 2, \cdots N$:
\beq
\label{xdxF}
 \frac{x^{j-1} dx}{ \sqrt{F_{2n}} } \;\;, \;\;\; j= 1 \sim n \;\;.
\eeq
 The differentials for $j=1 \sim n-1$ are the
 holomorphic differentials
  on this reduced surface and the last one $j=n$ has a pole
  at the infinities. This last one has been included due to the
 blow-up process and  physically implies  that the overall $U(1)$
  fails to decouple through this process.

  Finally let us examine eq. (\ref{st11}).
Let 
\beq
   \sum_{\ell}  \left( \frac{c_{\ell}}{\ell} \xi^{-\ell} \right)_{+}
  \equiv W_{K+1}^{\prime} (x)  \;,
\eeq
where $(\cdots)_{+}$ denotes  the part consisting of
 the non-negative powers of $\cdots$ in the Laurent expansion in $x$.
 We consider the case in which $W_{K+1}^{\prime} (x)$ is a
 polynomial  of degree
 $K$ $(K \geq n)$ in $x$ parameterized as
\beq
W_{K+1}^{\prime} (x)  \equiv \prod_{j=1}^{K}
 \left( x - \alpha_{j} \right) \;.
\eeq
  We  substitute eq. (\ref{xdxF}) for $d \omega_{i}$ in 
eq. (\ref{st11}) to obtain
\beq
\label{inftyres}
 0= res_{x=\infty} \left( W_{K+1}^{\prime}(x) \frac{x^{j-1}}
{\sqrt{F_{2n}}} dx \right) \;, \;\;\; j = 1 \sim n \;.
\eeq
Hence
\beq
 \frac{W_{K+1}^{\prime}(x)}{\sqrt{F_{2n}(x)} }
=  Q_{K-n}(x) + \sum_{\ell >n}  \frac{\beta_{\ell} }{x^{\ell}} \;\;,
\eeq
 where $Q_{K-n}(x)$ is a polynomial of degree $K-n$.
 We conclude
\beq
   F_{2n}(x) Q_{K-n}^{2}(x) = W_{K+1}^{\prime 2}(x)
  + f_{K-1}(x) \;\;,
\eeq
  where $f_{K-1}(x)$ is a polynomial  of degree $K-1$.
It still remains to be seen that  the function
$W_{K+1}(x)$ introduced above is in fact a  tree-level superpotential.
This is done in the next section.
  Let us also  note that our discussion
 suggests a family of superpotentials  continuously connected  when
   the kernel is more than one-dimensional.

\section{Classical Limit}

 In the classical limit, $\Lambda=0$ and
  the curve gets simplified largely: 
\beq
Y=z = \prod_{\ell =1}^{N} (x-p_{\ell}) \;\;.
\eeq
The original Seiberg-Witten differential becomes
\beq
\label{swclass}
 dS_{SW}^{(class)} =  x d \log z
   = \sum_{i=1}^{N} \frac{x}{x-p_{i}} dx \;\;.
\eeq
The period integrals over the $A_{i}$  cycles just pick up
 the poles at $p_{i}$:
\beq
  a_{i}^{(class)}  = p_{i} \;\;.
\eeq
The canonical holomorphic differentials take the following form:
\beq 
 d \omega_{i}^{(class)} = \frac{\partial}{\partial p_{i}}
  dS_{SW}^{(class)} \mid_{z} = \frac{dx}{x- p_{i}} \;\;.
\eeq
Eq.(\ref{dS}) is simply
\beq
 dS_{SW}^{(class)} = \sum_{i=1}^{N} p^{i} d \omega_{i} + N dx\;\;.
\eeq
 
  Let us  suppose that
\beq
z = \prod_{j =1}^{n} (x-\beta_{j})^{N_{j}}  \;\;,
 \;\;  \sum_{j=1}^{n} N_{j} = N \;\;, 
\eeq
  which means that $N_{j}$ poles of eq. (\ref{swclass})
 (or $N_{j}$ eigenvalues of the vev of the adjoint Higgs)  coalesce
 to one point $\beta_{j}$  for $j= 1 \sim n$ and
  $(N - n)$ non-intersecting  $A_{i}$ cycles vanish.  
The canonical holomorphic differentials  on this degenerate curve
  are
\beq
 d \omega_{j}^{(class, red)} =  \frac{dx}{x- \beta_{j}} \;\;, \;\;
 j = 1 \sim n \;\;,
\eeq
 so that
\beq
 a_{j}^{(red)} = \beta_{j}  \;\;.
\eeq
 Let us now examine the condition eq. (\ref{inftyres}).
  In the classical limit, eq. (\ref{inftyres}) becomes
\beq
 0= res_{x=\infty} \left( W_{K+1}^{\prime}(x) 
 d \omega_{j}^{(class, red)} \right) \;\;, \;\; j = 1 \sim n \;\;.
\eeq
  The residue is originally evaluated by the contour around
 infinity but by a contour deformation it becomes a residue
 at $x=\beta_{j}$.  We see that $\beta_{j}$ must coincide
 with  one of the roots $\alpha_{j}$ of  $W_{K+1}^{\prime}$.
 The vacuum values of the adjoint scalars are constrained to
 the extremum of $W_{K+1}$. We conclude that the function $W_{K+1}$
 is the tree-level superpotential as is promised.

\section{Discussion}

 Once the condition (\ref{kernel}) is imposed upon, we have a reduced
 curve of genus $g= n-1$.   Let us set $K=n$ for simplicity.
 In the context of the present paper, 
 the coefficients (denoted by $b_{\ell}$) of the polynomial
 $f_{K-1}$  are determined by the parameters $\alpha_{j}$
  of the superpotential.  In \cite{CIVCVTV, IM4, DVint}, 
  these $b_{\ell}$ as opposed to $u_{k}$ are treated as moduli.
The gluino condensate prepotential
  has been given  by the matrix model differential
  whose $b_{\ell}$ moduli derivatives  are almost holomorphic
 and these moduli derivatives are taken,
 keeping the coordinate $x$ fixed as opposed to $z$ fixed.
 In this setting the Whitham times have been identified with
  the symmetric polynomials made of $\alpha_{j}$.
Note that the $S_{i}$ moduli ( the $A_{i}$ cycle integral of the
  matrix model differential) as opposed to $a_{i}$  moduli
 have a quantum mechanical origin: they vanish as
 $\Lambda \rightarrow 0$. Also  
$ S\equiv \displaystyle{\sum_{i=1}^{n} S_{i}}$ is  non-vanishing and 
the cutoff must be introduced at the infinities of the
 curve.

\section{Acknowledgements}

The authors thank  E. Date, K. Ito, A. Morozov,  N. Sakai,
 K. Takasaki and Y. Yasui for helpful discussions and useful comments.
This work is supported in part by the Grant-in-Aid  for Scientific
 Research(14540264, 14540073) from the Ministry of Education,
Science and Culture, Japan.
 Supports from the 21st century COE programs at
  Osaka City University (H.I.) and at Nagoya University (H.K)
 are gratefully appreciated.
 
\newpage



\begin{thebibliography}{99}

\bibitem{SW}  
N.Seiberg and E.Witten, {\it Nucl.Phys.} {\bf B426}
(1994) 19 ( Erratum-ibid. {\bf B430} (1994) 485), hep-th/9407087.

\bibitem{SWC} 
A.Klemm, W.Lerche, S.Theisen and S.Yankielowicz,
{\it Phys.Lett.} {\bf B344} (1995) 169, hep-th/9411048; \\
P.Argyres and A.Farragi, {\it Phys.Rev.Lett.} {\bf 74} (1995) 3931,
hep-th/9411057; \\
A.Hanany and Y.Oz, {\it Nucl.Phys.} {\bf B452} (1995) 283,
hep-th/9505074.

\bibitem{SWINT}
A.Gorsky, I.Krichever, A.Marshakov, A.Mironov and A.Morozov,
{\it Phys.Lett.} {\bf B355} (1995) 466, hep-th/9505035;
E.Martinec and N.Warner, {\it Nucl.Phys.} {\bf B459} (1996) 97-112,
hep-th/9509161;
T.Nakatsu and K.Takasaki, {\it Mod.Phys.Lett.} {\bf A11} (1996)
157-168, hep-th/9509162;
R.Donagi and E.Witten, {\it Nucl.Phys.} {\bf B460} (1996) 299,
hep-th/9510101;
T.Eguchi and S.Yang, {\it Mod.Phys.Lett.} {\bf A11} (1996) 131-138,
hep-th/9510183;
H.Itoyama and A.Morozov, {\it Nucl.Phys.} {\bf B477} (1996)
855, hep-th/9511126; hep-th/9601168;
G. Bonelli and M. Matone, {\it Phys. Rev. Lett.} {\bf 77} (1996) 4712,
hep-th/9605090.

\bibitem{IM2}
H.Itoyama and A.Morozov, {\it Nucl.Phys.} {\bf B491} (1997) 529,
hep-th/9512161.

\bibitem{GMMM}
A.Gorsky, A.Marshakov, A.Mironov and A.Morozov,
{\it Nucl.Phys.} {\bf B527} (1998) 690-716, hep-th/9802007.

\bibitem{Edel}
J.Edelstein, M.Marino and J.Mas, {\it Nucl.Phys.} {\bf B541} (1999)
671, hep-th/9805172;
J.Edelstein, M. Gomez-Reino, M. Marino and J.Mas,
{\it Nucl.Phys.} {\bf B574} (2000) 587-619, hep-th/9911115.


\bibitem{whithamrev}
 K. Takasaki, {\it Prog. Theor. Phys. Suppl} {\bf 135} (1999) 53-74,
 hep-th/9905224; M. Marino 
{\it Prog. Theor. Phys. Suppl} {\bf 135} (1999) 29-52, hep-th/9905053;

\bibitem{DS}
 M. Douglas and S. Shenker, {\it Nucl.Phys.} {\bf B447} (1995)
271, hep-th/9503163;

\bibitem{LAG}
  L. Alvarez-Gaume, J. Distler, C. Kounnas and M. Marino,
{\it Int. J. Mod. Phys.} {\bf A11} (1996) 4745, hep-th/9604004;


\bibitem{CIVCVTV}
F.Cachazo, K.A.Intriligator and C.Vafa,
{\it Nucl.Phys.} {\bf B603} (2001) 3, hep-th/0103067;
F.Cachazo and C.Vafa, hep-th/0206017;
T.R. Taylor and C.Vafa,
{\it Phys. Lett.} {\bf B474} (2000) 130-137, hep-th/9912152.

\bibitem{DV}
R.Dijkgraaf and C.Vafa,
{\it Nucl.Phys.} {\bf B644} (2002) 3-20, hep-th/0206255;
{\it Nucl.Phys.} {\bf B644} (2002) 21-39, hep-th/0207106;
hep-th/0208048.

\bibitem{CDSW}
F.Cachazo, M.R.Douglas, N.Seiberg and E.Witten,
{\it JHEP} {\bf 0212} (2002) 071, hep-th/0211170;
F. Cachazo, N.Seiberg and E.Witten,
{\it  JHEP} {\bf 0302} (2003) 042, hep-th/0301006;
  hep-th/0303207.



\bibitem{ACDGN}
  J.R. David, E. Gava and K.S. Narian, hep-th/0304227;
  L.F. Alday, M. Cirafini, J.R. David, E. Gava and K.S. Narian,
 hep-th/0305217.

\bibitem{IK1}
 H. Itoyama, H. Kanno, 
{\it Phys. Lett} {\bf B573} (2003) 227-234, hep-th/0304184.

\bibitem{Konishi}
  K. Konishi, {\it Phys. Lett} {\bf B135} (1984) 439.

\bibitem{Argreview} 
 R. Argurio, G. Ferretti and R. Heise, hep-th/0311066.

\bibitem{DebHollo} 
  T.J. Hollowood, hep-th/0305023; R.Boels, J. de Boer, 
R. Duivenvoorden, and J. Wijnhout, hep-th/0304061;  hep-th/0305189.


\bibitem{RSFKRA}
See, for instance, H.M. Farkas and I. Kra, 
``Riemann Surfaces'',  Springer-Verlag 1991.

\bibitem{discriminants}
  S. Elitzur, A. Forge, A. Giveon, K. Intriligator and E. Rabinobicci,
  {\it Phys. Lett} {\bf B379} (1996) 121, hep-th/9603051;
  S. Terashima and S. K. Yang, {\it Nucl.Phys.} {\bf B519} (1998) 
453-469, hep-th/9706076.


\bibitem{Gopa}
R.Gopakumar,  {\it JHEP} {\bf 0305} (2003) 033, hep-th/0211100.

\bibitem{IM4}
 H.Itoyama and A.Morozov,
  {\it Nucl.Phys.} {\bf B657} (2003) 53-78,
hep-th/0211245.

\bibitem{DVint}
L.Chekhov and A.Mironov,
{\it Phys. Lett.} {\bf B552} (2003) 293-302, hep-th/0209085;
H.Itoyama and A.Morozov,
{\it Phys. Lett.} {\bf B555} (2003) 287-295,
hep-th/0211259;
L.Chekhov, A.Marshakov, A.Mironov and D.Vassiliev,
hep-th/0301071;
  H.Itoyama and A.Morozov,
{\it Prog.Theor.Phys.} {\bf 109} (2003) 433-463, hep-th/0212032;
M.Matone,
{\it Nucl.Phys.} {\bf  B656} (2003) 78-92, hep-th/0212253;
A. Dymarsky and V. Pestun, hep-th/0301135;
H. Itoyama and  A. Morozov, hep-th/0301136;
 S. Aoyama and T. Masuda, hep-th/0309232.

\end{thebibliography}
\end{document}